\documentclass[twocolumn]{aastex62}

\newcommand{\lat}{$Fermi$-LAT}

\hypersetup{linkcolor=red,citecolor=cyan,filecolor=blue,urlcolor=magenta}
\usepackage{subfigure}
\usepackage{graphicx}
\usepackage{epstopdf}
\shorttitle{Quasi-simultaneous infrared and $\gamma$-ray activities of CGRaBS J0733+0456}
\shortauthors{Liao et al.}

\begin{document}

\title{Multi-wavelength variability properties of CGRaBS J0733+0456: identifying a distant $\gamma$-ray blazar at {\it z} = 3.01}

\correspondingauthor{Neng-Hui Liao; Ting-Gui Wang}
\email{liaonh@pmo.ac.cn;twang@ustc.edu.cn}

\author[0000-0001-6614-3344]{Neng-Hui Liao}
\affiliation{Department of Physics and Astronomy, College of Physics, Guizhou University, Guiyang 550025, China}
\affiliation{Key Laboratory of Dark Matter and Space Astronomy, Purple Mountain Observatory, Chinese Academy of Sciences, Nanjing 210034, China}

\author[0000-0002-4757-8622]{Li-Ming Dou}
\affiliation{Center  for Astrophysics, School of Physics and Electronic Engineering, Guangzhou University,
Guang Zhou 510006, China}

\author[0000-0002-7152-3621]{Ning Jiang}
\affiliation{Key laboratory for Research in Galaxies and Cosmology, Department of Astronomy, The University of Science and Technology of China, Chinese Academy of Sciences, Hefei, Anhui 230026, China}

\author{Yi-Bo Wang}
\affiliation{Key laboratory for Research in Galaxies and Cosmology, Department of Astronomy, The University of Science and Technology of China, Chinese Academy of Sciences, Hefei, Anhui 230026, China}

\author[0000-0002-8966-6911]{Yi-Zhong Fan}
\affiliation{Key Laboratory of Dark Matter and Space Astronomy, Purple Mountain Observatory, Chinese Academy of Sciences, Nanjing 210034, China}

\author[0000-0002-1517-6792]{Ting-Gui Wang}
\affiliation{Key laboratory for Research in Galaxies and Cosmology, Department of Astronomy, The University of Science and Technology of China, Chinese Academy of Sciences, Hefei, Anhui 230026, China}

\begin{abstract}
We report on OVRO, {\it WISE}, {\it Swift} and {\it Fermi}-LAT observations of the high redshift blazar CGRaBS J0733+0456, from which significant flux variations in radio, infrared (IR) as well as $\gamma$-ray domains are detected. Particularly, the amplitude of long-term IR variation is over one order of magnitude and the IR variability timescale can be constrained as short as a few hours in the source frame. The IR and $\gamma$-ray light curves are found to be rather similar, and the strong quasi-simultaneous infrared and $\gamma$-ray flares are proved to be unique among the nearby $\gamma$-ray sources. This is the first time to identify a $\gamma-$ray blazar at redshift $z\geq 3$ with multi-wavelength flux variations (flares). Broadband spectral energy distributions in different flux states are constructed and theoretically described. The $\gamma-$ray flares from some blazars as distant as redshift $\sim$ 5 are expected to be detectable for {\it Fermi}-LAT.
\end{abstract}


\section{INTRODUCTION} \label{sec:intro}
Blazars, including flat-spectrum radio quasars (FSRQs) and BL Lacertae objects (BL Lacs), are an extreme subclass of Active Galactic Nuclei (AGNs). Their strong relativistic jets are beamed toward the observer and hence the highly variable non-thermal jet emissions are overwhelming \citep{BR78,Wagner1995,1997ARA&A..35..445U}. The luminous jet emission is characterized by a universal two-bump structure in log$\nu$F$\nu$-log$\nu$ plot, where one is believed to be synchrotron emission while the other one extending to the $\gamma$-ray domain is usually explained as inverse Compton (IC) scattering of soft photons from either inside (the synchrotron self-Compton, or SSC) and/or outside (external Compton, or EC) the jet by the same population of relativistic electrons generating the synchrotron emission \citep[e.g.,][]{1992ApJ...397L...5M,1993ApJ...416..458D,1994ApJ...421..153S,2000ApJ...545..107B}. Meanwhile, the detection of high energy neutrino event coincident with high energy $\gamma$-ray flares in TXS 0506+056 suggests that at least in some cases hadronic processes also take place \citep{2018Sci...361.1378I}.  Blazars at high redshifts, harboring supermassive black holes (SMBHs) heavier than one billion solar masses \citep{2004ApJ...610L...9R,2010MNRAS.405..387G}, are valuable targets since they shed lights on the formation and growth of the first generation of super massive black holes (SMBHs) and also on the evolution of AGN jets across different cosmic times \citep[e.g.,][]{2010A&ARv..18..279V}.

Benefiting from all-sky surveyors, like {\it Fermi}-LAT \citep{Atwood2009} and {\it WISE} \citep{Wright2010,Mainzer2014}, along with complementary observations from radio to X-rays, multi-wavelength campaigns become a regular approach to investigate the physical mechanism of blazars. Particularly, in addition to scanning in every a half year in mid-infrared (IR) bands at 3.4 $\mu$m (W1) and 4.6 $\mu$m (W2), {\it WISE} also provides about one dozen of exposures within 1-2 days in each epoch. Thus, it provides a chance for the mid-IR investigations in both long-term and intraday variability \citep{Jiang2012,Jiang2018,Mao2018}. Correlations between $\gamma$-ray emission of blazars and those in other windows of electromagnetic radiation have been frequently detected \citep[e.g.,][]{2010Natur.463..919A,2014ApJ...783...83L}. Meanwhile, considering the relatively limited angular resolution of the $\gamma$-ray observation, the correlated variations are essential to pin down the association between the $\gamma$-ray source and its low-energy counterpart \citep[e.g.,][]{2016ApJS..226...17L}. However, at the high redshift regime (i.e. $z\geq 3$), such crucial information are still lacking \citep{2017ApJ...837L...5A,2018ApJ...865L..17L}.

CGRaBS J0733+0456, first known as a radio emitter \citep{1995ApJS...97..347G} and identified as a high redshift quasar at $z$ = 3.01 \citep{2008ApJS..175...97H}, stands out for its large-amplitude mid-infrared variability during the systematical study of {\it WISE} flaring sources (Wang et al in preparation). Considering the significant radio variability \citep{2008ARep...52..278G} and the broadband flat radio spectrum spanning from 150~MHz to 15~GHz \citep{2011ApJS..194...29R,2017A&A...598A..78I}, it is not surprised that CGRaBS J0733+0456 is suggested as a blazar \citep{2009A&A...495..691M,2014ApJS..213....3M,2018ApJ...869..133F}. However, due to the low signal to noise ratio (S/N) of the W3 band detection listed in the ALLWISE catalog, the position of CGRaBS J0733+0456 in the ``WISE blazar strip'' \citep{2011ApJ...740L..48M} can not be well constrained. Interestingly, there is a $\gamma$-ray source cospatial with CGRaBS J0733+0456, named as FL8Y J0733.8+0455\footnote{https://fermi.gsfc.nasa.gov/ssc/data/access/lat/fl8y/} or 4FGL J0733.8+0455 in the fourth {\it Fermi}-LAT source catalog \citep{2019arXiv190210045T}.  In this Letter, multiwavelength data of CGRaBS J0733+0456 are collected and analyzed, and its broadband variability properties are reported (Section \ref{sec:data}), along with some discussions (Section \ref{sec:diss}). We adopt a ${\Lambda}$CDM cosmology with $\Omega_{M}$\,=\,0.3, $\Omega_{\Lambda}$\,=\,0.7, and a Hubble  constant of $H_{0}$\,=\,70\,km$^{-1}$\,s$^{-1}$\,Mpc$^{-1}$.

\section{DATA ANALYSIS AND RESULTS} \label{sec:data}
\subsection{{\it Fermi}-LAT Data}
We selected the first 10-year (MJD 54683$-$58335) {\tt SOURCE} type \lat~ {\tt Pass} 8 data ({\tt evclass} = 128 and {\tt evtype} = 3) with {the} energy range of $0.1-500$ GeV. The analysis was carried out with the updated software {\tt Fermitools} package of version {\tt 1.0.1}. The data filtration was accomplished with {\tt gtselect} and {\tt gtmktime} tasks, by adopting a maximum zenith angle of 90$\degr$ and ``{\tt DATA\_QUAL > 0}'' \&  ``{\tt LAT\_CONFIG==1}''. $\gamma$-ray flux and spectrum were extracted by the {\tt gtlike} task with {\tt unbinned} likelihood algorithm. The test statistic (TS = 2$\rm \Delta$log$\mathcal{L}$, \citealt{1996ApJ...461..396M}) was adopted to quantify the significance level of a $\gamma$-ray detection, where $\mathcal{L}$ represents the likelihood function, between models with and without the source. All 4FGL sources within 15$\degr$ of 4FGL J0733.8+0455 were considered. Parameters of background sources within 10$\degr$ of the target as well as two diffuse templates were left free, whereas others were frozen to be 4FGL values. After the {\tt gtlike} analysis, any new $\gamma$-ray sources (i.e TS $>$ 25) emerged in the subsequently generated TS residual map were added into the updated background model and the likelihood fitting was then re-performed. In the temporal analysis, we removed weak background sources (i.e TS $<$ 10) from the model. When TS value of one source is smaller than 10, the {\tt pyLikelihood UpperLimits} tool was adopted to calculate the 95\% confidential level (C. L.) upper limit instead.

First, the analysis of the entire 10-year LAT data suggests a new $\gamma$-ray source not included in 4FGL. Its TS value is 85 with an optimized $\gamma$-ray location of R.A. 118.49$\degr$ and DEC. 9.41$\degr$ (95\% C. L. error radius of 0.08$\degr$), likely associated with a flat spectrum radio source PMN J0753+0924 \citep{1995ApJS...97..347G}. After updating the background model, 4FGL J0733.8+0455 is proved as a significant (TS = 225) $\gamma$-ray source. Our analysis gives a 10-year averaged flux of (1.20 $\pm$ 0.15)$\times10^{-8}$ ph $\rm cm^{-2}$ $\rm s^{-1}$ with a pow-law index of 2.45$\pm$0.07 (i.e. $dN/dE \propto E^{-\Gamma}$, where $\Gamma$ is the photon index), in agreement with those reported in 4FGL. Following localization analysis confirms that CGRaBS J0733+0456 is the only known blazar candiate within the 95\% C. L. of $\gamma$-ray error radius.

\begin{figure}
\centering
\includegraphics[scale=0.8]{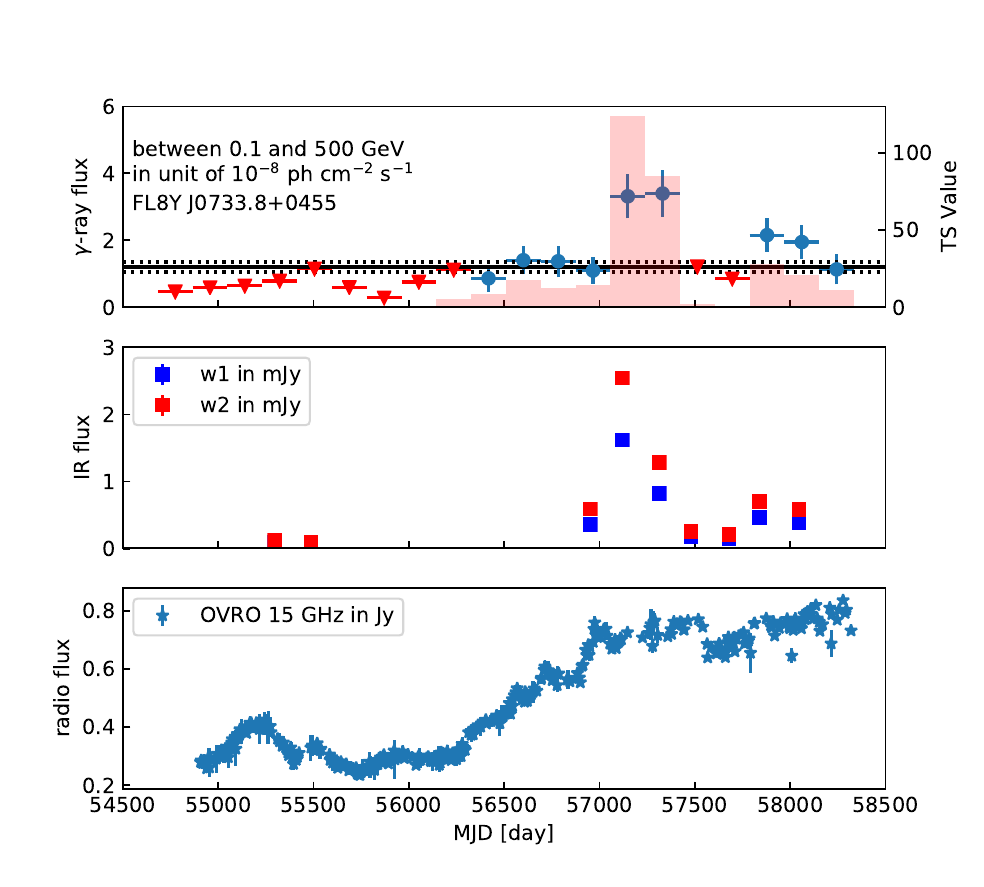}
\caption{Long-term multi-wavelength light curves of CGRaBS J0733+0456. Top panel: Half-year time bin $\gamma$-ray light curve, where the red triangles are the 95\% C.L. upper limits and the blue dots represent the flux data points. The red bars represent the TS values in each data bin. The 10-year averaged $\gamma$-ray flux (solid line) and its 1$\sigma$ uncertainty (dotted lines) are also marked. Middle panel: {\it WISE} IR light curves in W1 and W2 bands. Bottom panel: 15 GHz radio light curve from the OVRO project.}
\label{Fig.1}
\end{figure}

\begin{figure}
\centering
\includegraphics[scale=0.42]{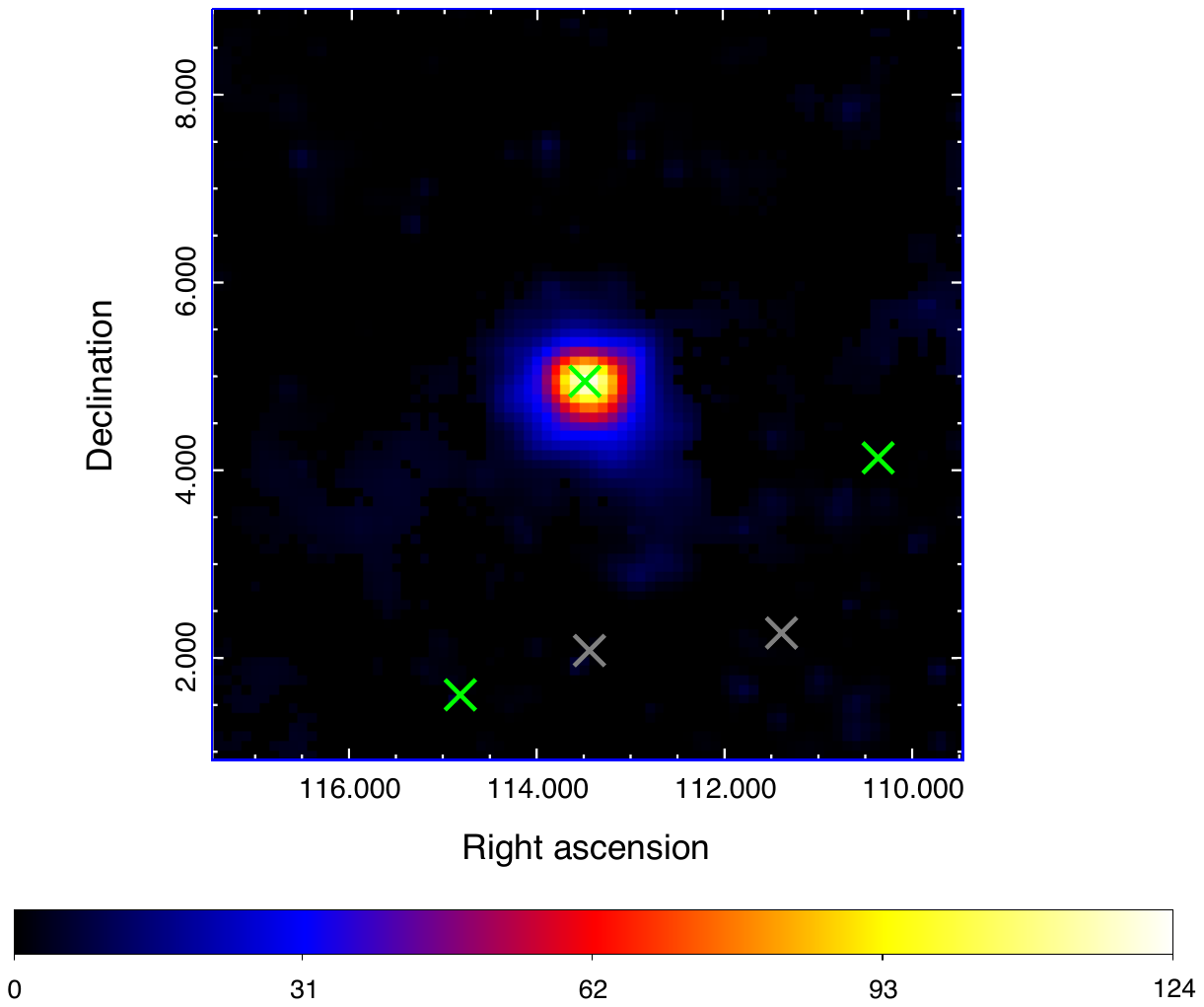}
\includegraphics[scale=0.42]{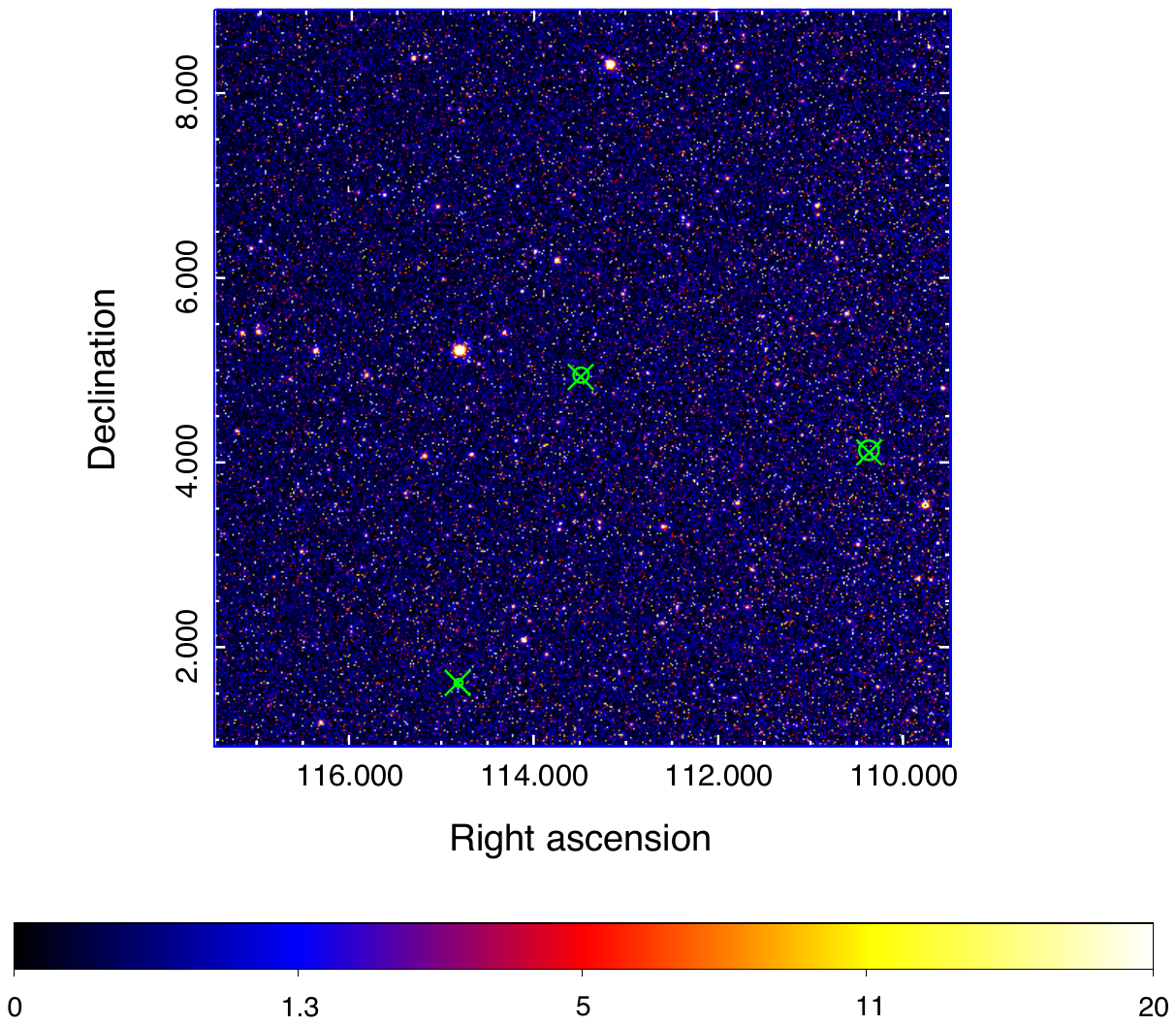}
\caption{8$\degr\times8 \degr$ views of the $\gamma$-ray and IR fields centered at CGRaBS J0733+0456. The left panel: the TS map of photons from 100~MeV to 500~GeV  corresponding to the flaring epoch, for which the diffuse backgrounds as well as 4FGL background sources are subtracted. The locations of the $\gamma$-ray sources are marked as cross symbols, in which the grey ones represent barely detected (TS $<$ 10) sources whereas the green ones correspond to significant $\gamma$-ray sources. The right panel: the NEOWISE W1 image in E4, in which circles are the 95\% C.L. error radii of the strong $\gamma$-ray sources and the crosses are the locations of their corresponding IR counterparts.}
\label{Fig.2}
\end{figure}

\begin{figure}
\centering
\includegraphics[scale=0.45]{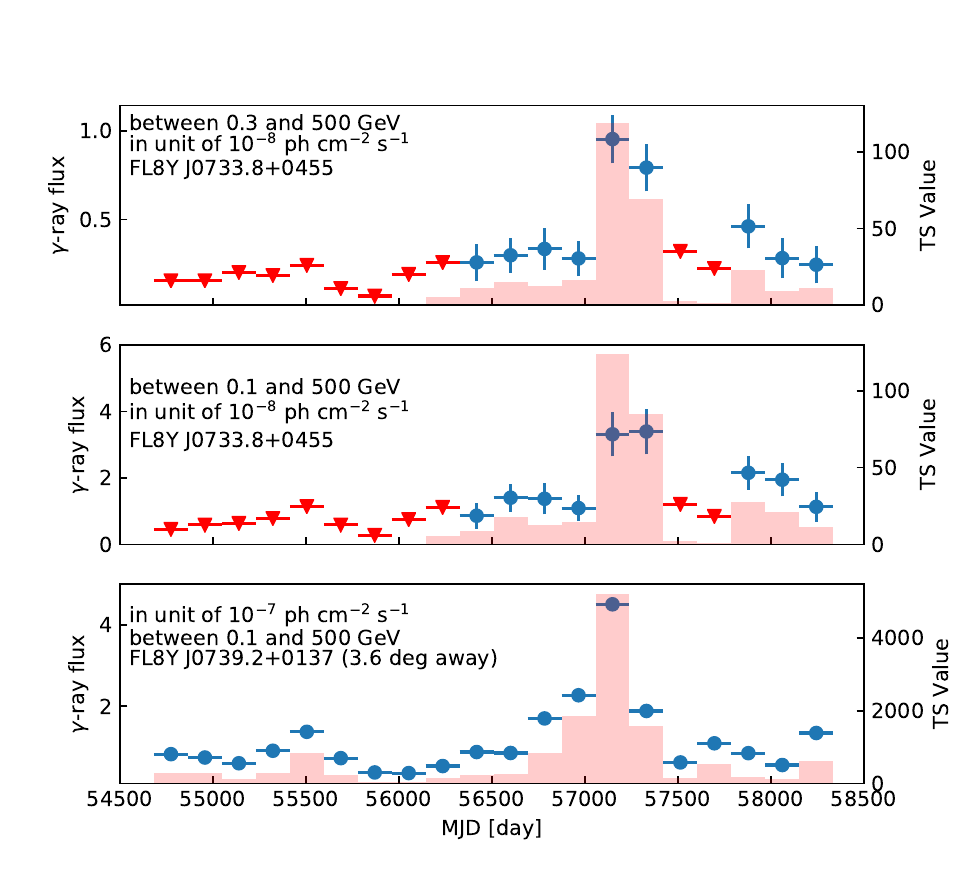}
\includegraphics[scale=0.45]{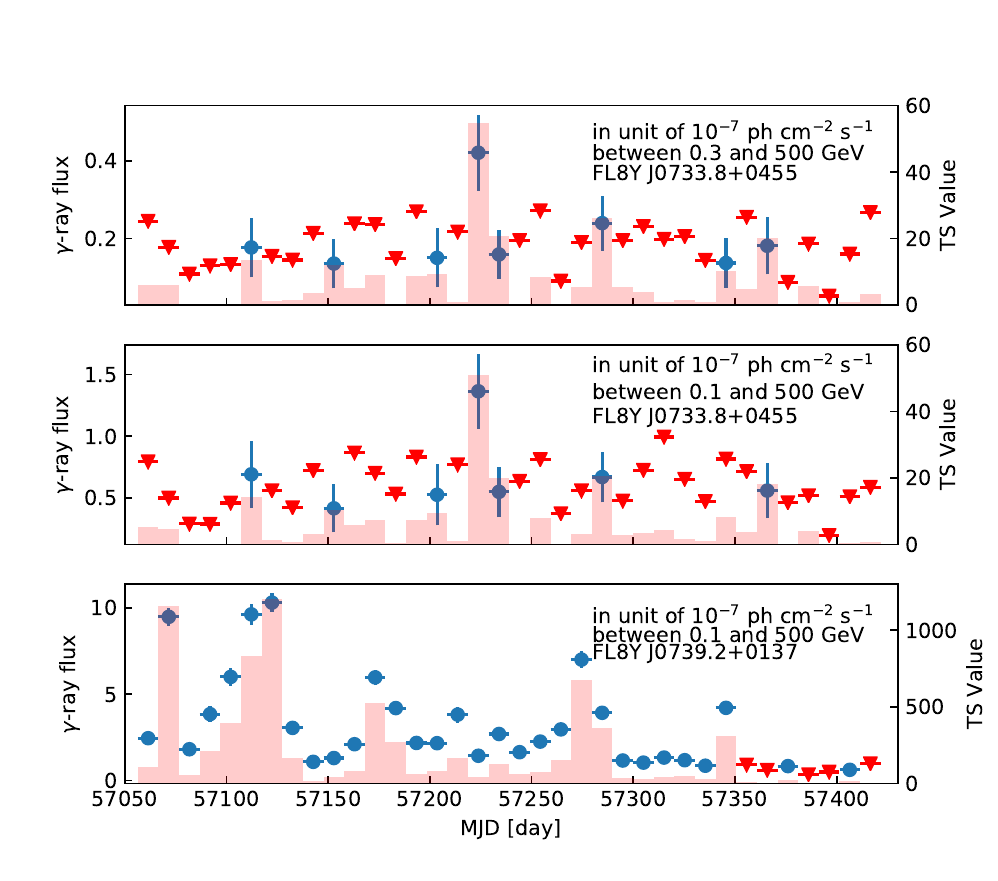}\\
\vspace{.3in}
\includegraphics[scale=0.45]{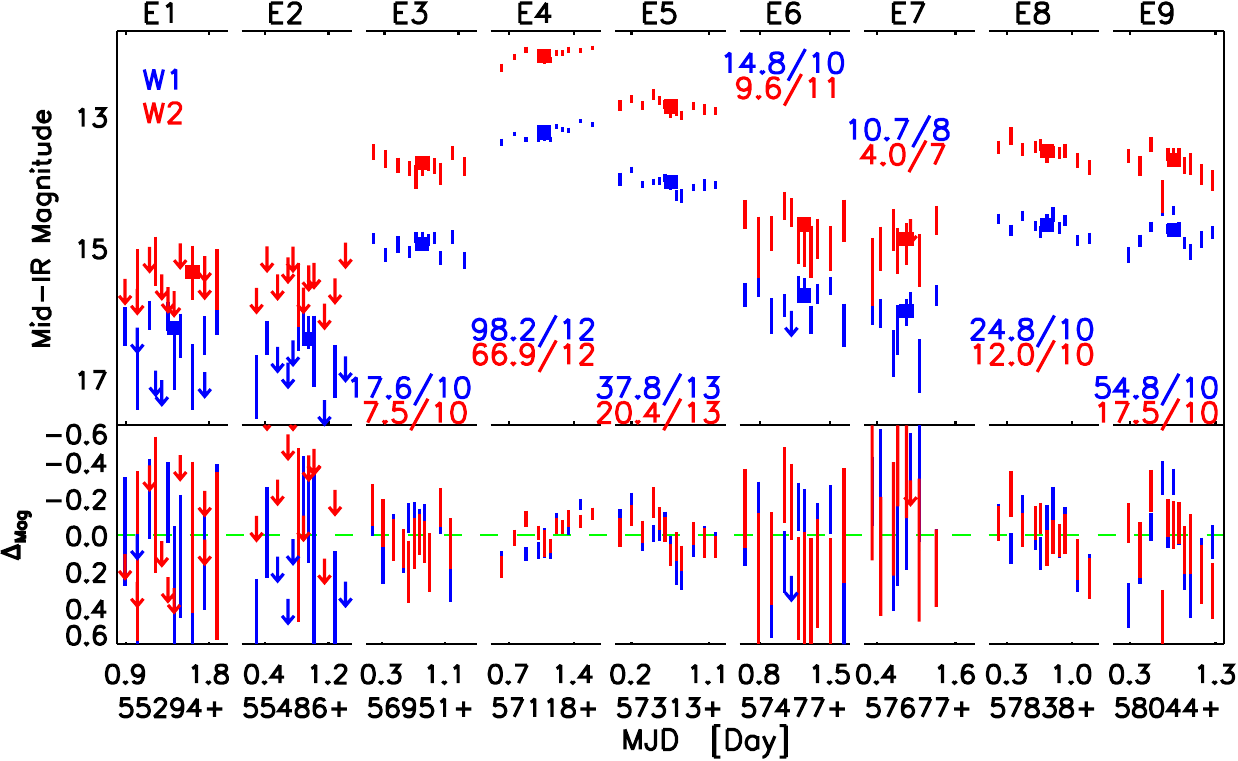}
\caption{Detailed $\gamma$-ray and IR light curves. {\bf Upper}: $\gamma$-ray light curves of FL8Y J0733.8+0455 (in different data energy ranges) as well as its neighbor FL8Y J0739.2+0137. Left panels: Half-year time bin light curves. Right panels: 10-day time bin light curves. The red triangles are the 95\% C.L. upper limits while the blue dots represent the flux data points. The red bars represent the TS values in each data bin. {\bf Lower}: 1st-row panel: The mid-IR light curves of CGRaBS J0733+0456. The fit results of $\chi^2/d.o.f$ to check the intraday variabilities for each epoch are listed in the bottoms. 2rd-row panel: The residual magnitudes after fit the light curves assuming a constant flux in each epoch. The square-symbols show the median values in each epoch. Note that data of E1, E2 are from ALLWISE while others are from NEOWISE.}
\label{Fig.3}
\end{figure}

We extracted a half-year time bin $\gamma$-ray light curve, see Figure \ref{Fig.1}. The target is undetectable (TS $<$ 4) for {\it Fermi}-LAT until Feb. 2013, which explains its absence in the third {\it Fermi}-LAT AGN catalog (3LAC, \citealt{2015ApJ...810...14A}). Since then, followed by a two years period of moderate flux state, a strong $\gamma$-ray flare appears, with peaking flux roughly three times of the 10-year averaged flux, see Figure \ref{Fig.2}. The target is quiescent after the flare, then a mild $\gamma$-ray activity occurs in 2017. In consideration of a bright nearby (3.6$\degr$ away) $\gamma$-ray source, 4FGL J0739.2+0137 (associated with PKS 0736+01), we also extracted its light curve to check its influence on the target light curve, see Figure \ref{Fig.3}. Though these two sources peaks at the same time bin, the strong flare of the target is still significant ($10\sigma$, which is calculated by $\rm TS_{var}$ defined in \citealt{2012ApJS..199...31N}) in $>$300~MeV (68\% C. L. containment angle of about 3$\degr$) light curve, suggesting that it is not artificial due to the flux variation of the neighbor. Furthermore, 10-day time bin $\gamma$-ray light curves were extracted. The separation of the peaking times of the two sources is about 100~days, which establishes the association of the strong flare with CGRaBS J0733+0456, see Figure \ref{Fig.3}. The 10-day peaking flux of CGRaBS J0733+0456 reaches (1.3 $\pm$ 0.4)$\times10^{-7}$ ph $\rm cm^{-2}$ $\rm s^{-1}$,  with a pow-law index of 2.10$\pm$0.17. The corresponding $\gamma$-ray luminosity is $(1.2\pm0.4) \times 10^{49}$ erg $\rm s^{-1}$. Although the limited statistics hampers us to search for short-term $\gamma-$ray variability, in contrast with the first 4.5-yr 95\% C. L. upper limit, $4\times10^{-9}$ ph $\rm cm^{-2}$ $\rm s^{-1}$, significant long-term $\gamma$-ray variability has been revealed.  Meanwhile, possible spectral hardening is found compared to the result from the 10-year averaged data.

\subsection{{\it Swift} Data}
The Neil Gehrels {\it Swift} Observatory \citep{2004ApJ...611.1005G} observed (ObsID: 00036786001) CGRaBS J0733+0456 in Dec. 2007. The XRT photon counting mode data and the UVOT UVW2 image were analyzed with the FTOOLS software version 6.22.1. For the XRT data, initially, the event cleaning with the {\tt xrtpipeline} using standard quality cuts was performed. Then the source spectra were extracted with {\tt xselect} from a circular region with a radius of 20 pixels while the background one from a larger circle (i.e. 50 pixels) in a blank area. To facilitate the spectral analysis, we produced the ancillary response file taken from the {\tt CALDB} database with {\tt xrtmkarf}. The grouped spectra were demanded to have at least 1 count per bin using the {\tt cstat} approach. Considering the 19 net photons, the absorption column density was set as the Galactic value (i.e. $\rm 7.5^{20}$ $\rm cm^{-2}$) and the spectral power-law index was frozen as 1.5 which is typical among high redshift blazars \citep[e.g.,][]{2013ApJ...763..109W},  giving an unabsorbed 0.5-10.0 keV flux of $\rm 1.29^{+0.86}_{-0.61}\times10^{-13}$ erg $\rm cm^{-2}$ $\rm s^{-1}$ ($C-Statistic/d.o.f$, 8.4/18; \citealt{1979ApJ...228..939C}). For the UVW2 image, an aperture photometry, using {\tt uvotsource} with a 5 arcsec circular aperture together with a background extraction in a larger source-free region, yielded no significant excess toward the target.

\subsection{{\it WISE} Data}
The mid-IR photometric data of CGRaBS J0733+0456 including a total of nine epochs (marked as E1..., E9) were collected from the ALLWISE and NEOWISE data release. The long-term light curves are presented in Figure\,\ref{Fig.1}. In order to check the long-term variations, the median magnitude values of W1 and W2 bands in each epoch were calculated\footnote{We only use the singe exposure photometry with S/N $>$3, and calculate the median values  following our previous works \citep[e.g.,][]{Dou2016,Jiang2016}}. The fluxes are in low state during the first two epochs in 2010. There is a giant mid-IR flare in 2015, the flux peaks in E4 epoch (MJD$\sim$57118, 2015-4-6), which W1 (W2) band photometry gives 13.19 (12.07) magnitude. It is 3.0 (3.5) magnitude in W1 (W2) band brighter than the low-flux state in E1 and E2 epochs.

In short-term, among the $\sim$10 exposures within $\sim$1 day in each epoch, only the best-quality single-frame images (`qual$\_frame$'\,$>$\, 0.5) were selected, see Figure \ref{Fig.3}.  In most of the single exposures in the E1 and E2 epochs, only upper limits were given due to the limited S/Ns. We adopted the $\chi^2$-statistic to search any short-term variations in other epochs with a null hypothesis of a constant flux.  Significant intraday variations in W1 band were detected in E4, E5, E8 and E9 epochs (reduced $\chi^2\,>\,2$). Meanwhile, intraday variations were also detected in W2 band in E4, E5 and E9 epochs. The cross correlations between W1 and W2 band in E4, E5 and E9 epochs were then calculated.  Significant cross correlations were detected in E4 and E5 epoch, with the cross correlation factors of 0.73 and 0.65, respectively. No significant cross correlation was detected in E9 epoch. The intraday variability properties in those epochs were investigated, following our previous works \citep[e.g.,][]{Ai2010, Jiang2012}. The W1 band variability amplitudes are 0.09, 0.08, 0.10 and 0.20 magnitude, respectively, in E4, E5, E8 and E9 epochs. The W2 band variability amplitudes are 0.09, 0.05, and 0.16 magnitude, respectively, in E4, E5, and E9 epochs. The confidence levels of variability significance in W1 band are 5.0, 3.0, 2.4, and 4.2$\sigma$, respectively, in E4, E5, E8 and E9 epochs. While in W2 band they are 4.3,  2.4, and 2.4$\sigma$, respectively,  in E4, E5, and E9 epochs. The power spectrum density in E4 was also calculated, from which the timescale of the intraday variability is 3$\sim$4 hours in rest frame. A intraday bluer-when-brighter trend was detected in each epoch, though the trend is tentative considering the uncertainties. Such a trend disappears among different epochs (i.e. from E3 to E9). The average value of the median W1-W2 colors in epoch E3 to E9  is 1.09, while the one is $\sim$0.75 in epoch E1 to E2. The IR spectral variability behavior of CGRaBS J0733+0456 is consistent with those of other blazars \citep[e.g.,][]{2017ApJ...834..113M,Jiang2018}.

\subsection{OVRO light curve}
CGRaBS J0733+0456 is included in the Owens Valley Radio Observatory (OVRO) 40 m telescope monitoring program\footnote{http://www.astro.caltech.edu/ovroblazars/}. This program begins in 2007 and encompasses over 1800 objects that are mainly from the CGRaBS survey \citep{2008ApJS..175...97H}. The radio light curves are well sampled with observations for each source occurring twice per week, at a frequency of 15 GHz \citep{2011ApJS..194...29R}. Significant radio variability of CGRaBS J0733+0456 displays in the OVRO light curve, see Figure \ref{Fig.1}. It maintains in a low radio flux state ($\lesssim$ 0.4 Jy) before 2013. Since then the flux is doubled after a two years ascent phase and keeps in a high radio flux state until now.

\subsection{Implications of multi-wavelength variability}
Significant flux variations in radio, mid-IR (shifted to near-IR in the source frame) as well as $\gamma$ rays have been detected for CGRaBS J0733+0456. Especially, IR emissions at epoch E4 is over 15 times of those at epoch E1, strongly suggesting that the jet emission is dominated then. Moreover, the IR and $\gamma$-ray light curves are rather similar, including a quiescent period at early time of {\it Fermi}-LAT operation, a strong flare event in 2015 as well as a mild activity in 2017. For the $\gamma$-ray flare of CGRaBS J0733+0456 alone, there could be possible contamination from PKS 0736+01. However, the quasi-simultaneous IR flare detected in CGRaBS J0733+0456 does not emerge for PKS 0736+01, together with the well angular resolution of the WISE observation (see Figure \ref{Fig.2}), strongly supporting the validness of the association between CGRaBS J0733+0456 and 4FGL J0733.8+0455. The IR and $\gamma$-ray fluxes of CGRaBS J0733+0456 in the flaring state are checked and it is is consistent with other blazars in the perspective of infrared-$\gamma$-ray connection \citep{2016ApJ...827...67M}. In fact, a tight connection between emissions from low-energy frequencies (i.e. optical and near-IR bands) and $\gamma$ rays of FSRQs are widely accepted thanks to the detections of simultaneous flares at the corresponding wavelengths \citep[e.g.,][]{2012ApJ...756...13B} and significant variation of optical polarization properties during a giant $\gamma$-ray flare \citep[e.g.,][]{2010Natur.463..919A}. Therefore, cospatiality of these two emissions are suggested and the leptonic processes likely play an important role in generating of $\gamma$-ray emission, which are further supported by CGRaBS J0733+0456. By contrast, the radio variation is moderate since it is likely from a more extended region than the near-IR and $\gamma$-ray bands. Nevertheless, the beginning of the radio ascent phase is coincident with ``emergence" of the $\gamma$-ray emission. Meanwhile, the radio emission maintains in the high flux state up to now because the corresponding electrons cooled much slowly than that generate the near-IR and $\gamma-$ray flares. Based on the consistent multiwavelength light curves, we conclude that CGRaBS J0733+0456 is the low-energy counterpart of FL8Y J0733.8+0455.

In short-term, intraday IR variability is identified, in which the variability timescale is constrained as short as 4 hours. Therefore a compact radiation region is inferred,
\begin{equation}
R_{j}^{\prime}\lesssim\delta ct_{\rm var,IR}\sim 10^{16} \rm cm \frac{\delta}{30} \frac{t_{\rm var,IR}}{\rm 4~hrs},
\end{equation}
where $\delta$ is the Doppler factor, c is speed of the light. Such a deduction is also supported by the detections of fast $\gamma$-ray variability of FSRQs \citep[e.g.,][]{1995MNRAS.273..583D,2008MNRAS.384L..19B,2011A&A...530A..77F,2018Galax...6...68L}. Sometimes the $\gamma$-ray variability timescale could be down to a few minutes \citep{2011ApJ...730L...8A,2016ApJ...824L..20A}. Assuming near-IR and $\gamma$-ray photons of CGRaBS J0733+0456 are from the same region, to successfully escape from heavy absorption on $\gamma$ rays via $\gamma\gamma$ process, the Doppler factor should be high enough. The highest energy of the detected $\gamma$-ray photon is about 10~GeV while the absorption soft photons could be from the jet itself and the accretion system externally. In the former case, if we set the observed luminosity of soft photons (at a few keV) as $5\times10^{46}$ erg ${\rm s^{-1}}$ since there is no simultaneous X-ray observations here, and the radius of the $\gamma$-ray radiation region is as same as the absorption length, $\delta \gtrsim$ 10 is given. Alternatively, little information of the external absorption photons at several tens of eVs is acquired, which prevents us to set a reliable constraint on the Doppler factor.

\section{DISCUSSION and SUMMARY} \label{sec:diss}
There are only two $\gamma$-ray sources marked as blazars beyond redshift 3 in the 3LAC \citep{2015ApJ...810...14A}. Fortunately, the number of such sources is increasing. Recently, detections of $\gamma$-ray emission of five new blazars with $z >$ 3.1 have been claimed, with the highest redshift of 4.3 \citep{2017ApJ...837L...5A}. Additional candidates/sources include GB6 B1427+5149 at $z=3.03$ \citep{2018arXiv181107961M} and B3 1428+422 at $z=4.72$ \citep{2018ApJ...865L..17L}, the farthest object among the 105 month {\it Swify}-BAT all sky hard X-ray survey \citep{2018ApJS..235....4O}. However, identifications of these high redshift $\gamma$-ray blazar candidates are based on the cospatiality between the $\gamma$-ray sources and the low-energy counterparts. \cite{2017ApJ...837L...5A} presents broadband spectral energy distributions (SEDs) of their five sources and argues that they share a typical FSRQ shape, but majority of their multiwavelength data are un-simultaneous. Though strong optical and $\gamma$-ray flares of NVSS J163547+362930 ($z$ = 3.6) are identified \citep{2018ApJ...853..159L}, these flares appear at different times. The successful detection of the simultaneous IR and $\gamma-$ray flares ranks CGRaBS J0733+0456 as the first unambiguously identified $\gamma$-ray blazar beyond redshift 3. The broadband activities of CGRaBS J0733+0456 also confirm violent behaviors detected from other high redshift blazars \citep[e.g.,][]{2013A&A...556A..71A,2014MNRAS.444.3040O,2015ApJ...799..143A,2018ApJ...853..159L}. Though high redshift blazars tend to be weak and spectrally soft $\gamma$-ray sources, the detection prospect can be significantly enhanced in the presence of the energetic flares.

\begin{figure}
\centering
\includegraphics[scale=0.99]{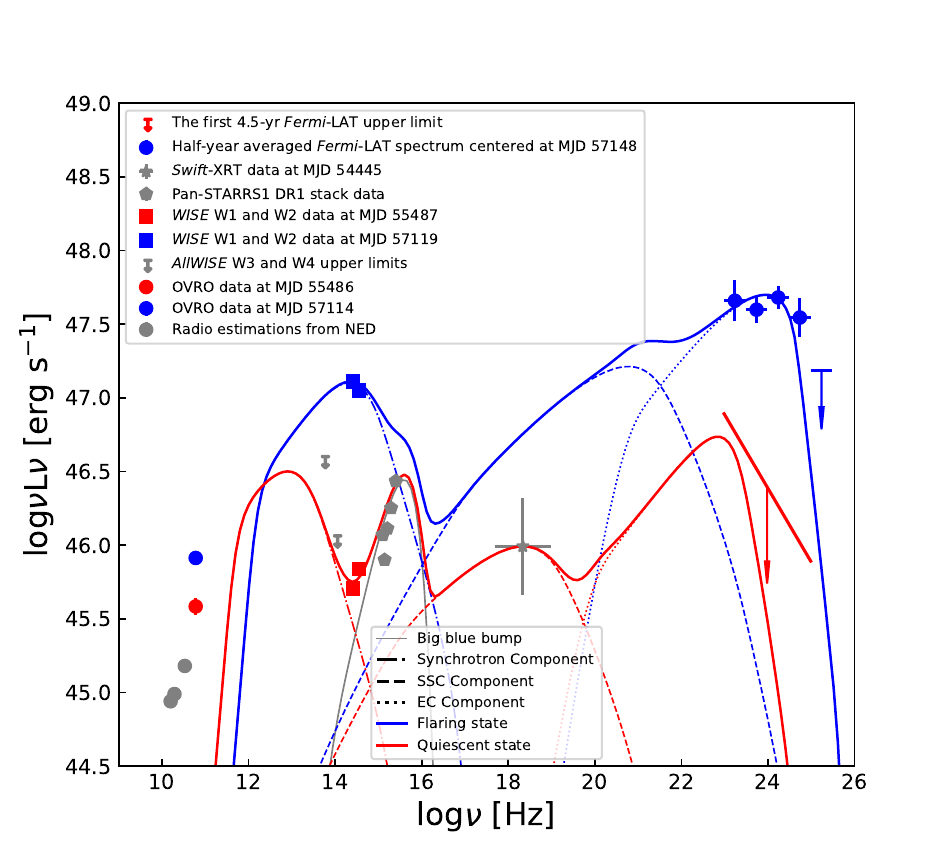}
\caption{SEDs in different flux states of CGRaBS J0733+0456 along with the theoretical descriptions. The color grey data points are un-simultaneous data and the grey line represents the accretion disk emission with $L_{d} = 6\times 10^{46}$ erg $\rm s^{-1}$ ($\sim 0.5L_{edd}$) and $M_{BH} = 5\times 10^{8}M_{\odot}$. The color blue and red data points and lines correspond to the flaring and the quiescent states, respectively.}
\label{sed}
\end{figure}

Broadband SEDs of CGRaBS J0733+0456 are shown in Figure \ref{sed}. Un-simultaneous multi-wavelength data include radio flux densities obtained from NED, stack five-band magnitudes from the first data release from the Panoramic Survey Telescope and Rapid Response System (Pan-STARRS, \citealt{2016arXiv161205560C,2016arXiv161205243F}) as well as the X-ray estimation from {\it Swift}-XRT analyzed here. Meanwhile, the OVRO data at MJD 55486, {\it WISE} observations at MJD 55487 and the first 4.5-yr {\it Fermi}-LAT 95 C.L. upper limit represent the low flux state SED, while OVRO data at MJD 57114, {\it WISE} observations at MJD 57119 and a half-yr averaged {\it Fermi}-LAT $\gamma$-ray spectrum centered at MJD 57148 correspond to the high flux state SED. In addition to the remarkable IR variability amplitude, the IR spectra are also variable, indicating that the synchrotron bump peaks at higher frequencies during the flaring epoch. The Pan-STARRS and the E2 {\it WISE} data are extracted by the emission from a standard \cite{1973A&A....24..337S} disk. The accretion disk extends from 3$R_{s}$ to 2000$R_{s}$, where $R_{s}$ is the Schwarzschild radius, and produces a total luminosity $L_{d}=\eta\dot{M}c^{2}$ in which $\dot{M}$ is the accretion rate and the accretion efficiency $\eta$ is set as a typical value, 0.1. The accretion disk emission exhibits a multi-temperature radial profile, and the local temperature at a certain radius $R_{d}$ is,
\begin{equation}
T^{4} = \frac{3R_{s}L_{d}}{16\pi\eta\sigma_{MB}R_{d}^{3}}\left[1-(\frac{3R_{s}}{R_{d}})^{1/2}\right].
\end{equation}
Meanwhile, a simple homogeneous one-zone leptonic scenario is adopted to describe the jet emission which is from a relativistic compact blob with a radius of $R_{j}^{\prime}$ embedded in the magnetic field. The emitting electrons follow a broken power-law distribution,
\begin{equation}
N(\gamma )=\left\{ \begin{array}{ll}
                    K\gamma ^{-p_1}  &  \mbox{ $\gamma_{\rm min}\leq \gamma \leq \gamma_{br}$} \\
            K\gamma _{\rm br}^{p_2-p_1} \gamma ^{-p_2}  &  \mbox{ $\gamma _{\rm br}<\gamma\leq\gamma_{\rm max}$,}
           \end{array}
       \right.
\label{Ngamma}
\end{equation}
where $\rm \gamma_{br}$ is the electron break energy, $\rm \gamma _{min}$ and $\rm \gamma _{max}$ are the minimum and maximum energies of the electrons, $K$ is the normalization of the particle number density, and the $p_{1,2}$ are indices of the broken power-law particle distribution. Both synchrotron and IC processes are considered, along with the synchrotron self-absorption process and the Klein$-$Nishina effect in the IC scattering. Assuming a conical jet geometry, the distance between the jet radiation region and the central SMBH $\rm r_{diss}$ = $\delta R_{j}^{\prime}\sim$ 0.1~pc. Based on the monochrome UV flux at 1350~\AA~from the optical spectrum of CGRaBS J0733+0456 \citep{2008ApJS..175...97H}, the scale of the broad line region (BLR) can be inferred as $\rm r_{BLR} \sim$ 0.3~pc \citep{2015ApJ...801....8K}. Since the strong Ly$\alpha$ line is detected, Ly$\alpha$ line emission is adopted as the external soft photons in the EC process. Meanwhile, the energy density of the external soft photons can be also estimated from the optical spectrum of CGRaBS J0733+0456 \citep{2008ApJS..175...97H}, $\rm U_{ext} = L_{Ly\alpha}/4\pi r_{BLR}^{2}c \simeq 9.4\times10^{-4}$ erg $\rm cm^{-3}$. The transformations of frequency and luminosity between the jet frame and the observational frame are $\nu = \delta\nu^{\prime}/(1+z)$ and $\nu L_{\nu} = \delta^{4}\nu^{\prime}L^{\prime}_{\nu^{\prime}}$. The simple leptonic model can provide reasonable descriptions of both SEDs (see Figure \ref{sed}), though there are only a little information for the low flux state. The input parameters of the jet radiation models are listed in Table \ref{tpara}. A significant enhance of the Doppler factor as well as the $\rm \gamma_{br}$ due to ejecta of a new jet blob could account for the multi-wavelength flares of CGRaBS J0733+0456. \cite{2011MNRAS.411..901G} performs a systematical SED modeling study of 19 high redshift blazars with one-year averaged {\it Fermi}-LAT data. Their typical magnetic field intensity is $\sim$ 1 Gauss and bulk Lorentz factor of the jet blob is $\sim$ 13. Meanwhile, \cite{2018ApJ...856..105A} revisits the SED of QSO J0906+6930, the farthest known blazar so far, though it has not been identified as a $\gamma$-ray emitter yet. To explain the $\gamma$-ray ``off"  SED, a bound 6$< \delta <$ 11.5 is set. In brief, the input parameters of the jet radiation models here are in agreement with SED modeling studies for other high redshift blazars.

In summary, we present a broadband temporal view of the high redshift blazar CGRaBS J0733+0456. Significant radio, IR and $\gamma$-ray flux variations are identified. In particular, violent long-term IR variability with amplitude over one order of magnitude and intraday IR variations with variability timescale down to a few hours have been detected. Moreover, the IR and $\gamma$-ray light curves share a rather similar shape, providing a decisive proof of the association between CGRaBS J0733+0456 and FL8Y J0733.8+0455. Broadband quasi-simultaneous SEDs correspond to different flux states can be well reproduced by the homogeneous one-zone leptonic radiation scenario. If similar strong activities occur in blazar candiates with $z \gtrsim$ 5 \citep{2004ApJ...610L...9R,2012MNRAS.426L..91S,2014MNRAS.440L.111G,2014ApJ...795L..29Y}, detection of their $\gamma$-ray emissions are anticipated.

\acknowledgments
We appreciate the instructive suggestions from the anonymous referee that led to a substantial improvement of this work. We also appreciate Prof. Roger W. Romani for sharing their CGRaBS spectrum of CGRaBS J0733+0456. This research has made use of data obtained from the High Energy Astrophysics Science Archive Research Center (HEASARC), provided by $\rm NASA^{\prime}$s Goddard Space Flight Center. This research makes use of data products from the Wide-field Infrared Survey Explorer, which is a joint project of the University of California, Los Angeles, and the Jet Propulsion Laboratory/California Institute of Technology, funded by the National Aeronautics and Space Administration. This research also makes use of data products from NEOWISE-R, which is a project of the Jet Propulsion Laboratory/California Institute of Technology, funded by the Planetary Science Division of the National Aeronautics and Space Administration. This research has made use of data from the OVRO 40-m monitoring program (Richards, J. L. et al. 2011, ApJS, 194, 29) which is supported in part by NASA grants NNX08AW31G, NNX11A043G, and NNX14AQ89G and NSF grants AST-0808050 and AST-1109911. The Pan-STARRS1 Surveys (PS1) and the PS1 public science archive have been made possible through contributions by the Institute for Astronomy, the University of Hawaii, the Pan-STARRS Project Office, the Max-Planck Society and its participating institutes, the Max Planck Institute for Astronomy, Heidelberg and the Max Planck Institute for Extraterrestrial Physics, Garching, The Johns Hopkins University, Durham University, the University of Edinburgh, the Queen's University Belfast, the Harvard-Smithsonian Center for Astrophysics, the Las Cumbres Observatory Global Telescope Network Incorporated, the National Central University of Taiwan, the Space Telescope Science Institute, the National Aeronautics and Space Administration under Grant No. NNX08AR22G issued through the Planetary Science Division of the NASA Science Mission Directorate, the National Science Foundation Grant No. AST-1238877, the University of Maryland, Eotvos Lorand University (ELTE), the Los Alamos National Laboratory, and the Gordon and Betty Moore Foundation.

This work was supported in part by the NSFC under grants 11525313 (i.e., Funds for Distinguished Young Scholars), 11833007, 11421303, 11733001 and 11703093, and National Basic Research Program of China (grant No. 2015CB857005) as well as U1631109 and U1731104, jointly supported by CAS and NSFC.

\vspace{5mm}
\facilities{{\it Fermi} (LAT); {\it WISE}; {\it Swift}}; OVRO

\clearpage

\begin{deluxetable}{lccccccccc}
\scriptsize
\tablenum{1} \tablewidth{0pt}
\tablecaption{List of parameters used to construct the theoretical jet SEDs in Figure \ref{sed}}
\tablehead{ \colhead{Model} &\colhead{$p_{1}$} &\colhead{$p_{2}$} &\colhead{$\gamma_{br}$} &\colhead{$K$[$\rm cm^{-3}$]} &\colhead{$B$[Gauss]} &\colhead{$\delta$} &\colhead{$R_{j}^{\prime}$[cm]}}
\startdata
Flaring state &2.2 &5.0 &1826 &$\rm 3.4\times10^{5}$ &1 &26 &$\rm 1.1\times10^{16}$ \\[3pt]
Quiescent state &2.2 &5.0 &570 &$\rm 3.0\times10^{4}$ &1 &10 &$\rm 7.8\times10^{16}$ \\[3pt]
\enddata
\tablecomments{$p_{1,2}$ are the indexes of the broken power-law radiative electron distribution; $\gamma_{br}$ is the break energy of the electron distribution; $K$ is the normalization of the particle number density; $B$ is the magnetic field strength; $\delta$ is the Doppler boosting factor and $R_{j}^{\prime}$ is the radius of the emission blob in the jet comoving frame. The minimum and maximum energies of the electrons are set as 100 and 10 times of the $\gamma_{br}$, respectively. The energy density of the Ly$\alpha$ line emission emission is estimated as $9.4\times10^{-4}$ erg $\rm cm^{-3}$ in the rest frame. \tiny}
\label{tpara}
\end{deluxetable}

\end{document}